\documentclass[aps,pre,showpacs,groupedaddress,showkeys]{revtex4-1}  % for review and submission
\usepackage{graphicx}  % needed for figures
\usepackage[caption=false]{subfig}
\usepackage{dcolumn}   % needed for some tables
\usepackage{bm}        % for math
\usepackage{amssymb}   % for math
\usepackage{amsmath}
\usepackage{epstopdf}

\begin{document}
\title{Exact solutions to the (1+1)-dimensional nonlinear Maxwell equations in the orthogonal curvilinear coordinates}

\author{Liang Hu}
\author{Xiao Zhang}
\author{Dazhi Zhao }
 \email{zhaodazhimail@163.com}
\author{MaoKang Luo}
 \email{makaluo@scu.edu.cn}

\affiliation{Department of Mathematics, Sichuan University 610065, Chengdu, Sichuan, China}

\date{\today}

\begin{abstract}
Characterizing electromagnetic wave propagation in nonlinear and inhomogeneous media is of great interest from both theoretical and practical perspectives, even though it is extremely complicated. In fact, it is still an unresolved issue to find the exact solutions to the nonlinear waves in the orthogonal curvilinear coordinates. In this paper, we present an analytic method to handle the problem of electromagnetic waves propagation in arbitrarily nonlinear and particularly inhomogeneous media without dispersion. Through the exact solutions of the (1+1)-dimensional nonlinear Maxwell equations, we discuss some nonlinear phenomena, including cylindrical shock waves, free nonlinear oscillations, and nonlinear superposition of waves.
\end{abstract}

%\linenumbers

\maketitle
\section{Introduction}
Along with the discovery of media possessing nonlinear electromagnetic properties, understanding electromagnetic wave propagation in nonlinear media has become a topical problem in physics \cite{1974G.B.Whitham,1984Shen,2008Boyd,1982Bloembergen685}. Given the fact that most of media in nature or artificial metamaterials are nonlinear and inhomogeneous, it is of special importance to study the behavior of electromagnetic fields in media with these two properties. However, this problem is yet to be solved \cite{2010Petrov190404,2010Xiong57602}, especially for exact solutions which are essential in understanding the related physical processes and in developing new computational asymptotic methods \cite{2014Polyanin409,2013Polyanin115,2013Polyanin77,2014Polyanin16,2008Popovych209,2002Zamboni-Rached217,
PhysRevE.64.066603,2006Zamboni-Rached1804,2015Ambrosio2584}.
Traditional methods like the inverse scattering method and B\"{a}cklund transformation are efficient for constructing exaction solutions in the theory of plane waves \cite{2006HE1141,1999He699,2006He700}, but they turn out to be inapplicable in cylindrical and spherical waves \cite{2010Kudrin537}. Meanwhile, finding new and physically meaningful exact solution of nonlinear waves is a highly urgent task, due to its own limitation of coupled-wave equation method \cite{2015Xiong11071} and difficulty to acquire the satisfactory description of strongly nonlinear phenomenons (i.e. self-steepening and shock waves) by perturbation theory.

Notably, Ref. \cite{2010Petrov190404} reports that they can obtain an exact solution to the cylindrical electromagnetic wave propagation problem in a medium which has a special form of nonlinearity - with the dielectric function being an exponential function of electric field. Following their lead, a number of related works have been reported recently \cite{2010Xiong57602,2010Kudrin537,2015Xiong11071,2011Eskin67602,2011Xiong43841,2011Xiong63845,2012Petrov55202,
2012Xiong16602,2012Xiong16606,2013Chen35202,2016Ranjbar19}. To the best of our knowledge, however, serious difficulties have been met in studying other nonlinear forms.

Here, we propose a method to exactly solve the (1+1)-dimensional Maxwell equations in arbitrarily nonlinear and particularly inhomogeneous media without dispersion. We show that if exact solutions of a linear system are given, then exact solutions of the corresponding two nonlinear systems can also be obtained;  and the exact solutions of a nonlinear system can be got from either of the corresponding two linear systems. By usage of this connection, we can investigate some nonlinear phenomena by analytic methods. As illustrations, we analyze cylindrical shock waves and nonlinear oscillations of cylindrical waves. We also show how to deal with two waves propagation in a nonlinear medium.

\section{Electromagnetic model}
Consider electromagnetic fields in nonmagnetic and nondispersive media in the system of orthogonal curvilinear coordinates $(u,v,w)$. If the fields are independent of $v$ and $w$, and the polarization direction of the electromagnetic waves is along the $v$ axis, viz. $v$-axis polarization, then the (1+1)-dimensional Maxwell equations can be written as
\begin{equation*}
\begin{aligned}
\frac{1}{l_u}\frac{\partial E_v(u,t)}{\partial u}+\frac{1}{l_ul_v}\frac{\partial l_v}{\partial u}E_v(u,t)&=-\mu_0\frac{\partial H_w(u,t)}{\partial t},
\\
-\frac{1}{l_u}\frac{\partial H_w(u,t)}{\partial u}-\frac{1}{l_ul_w}\frac{\partial l_w}{\partial u}H_w(u,t)&=\epsilon(u,E_v)\frac{\partial E_v(u,t)}{\partial t},
\end{aligned}
\end{equation*}
where $\epsilon(u,E_v)=dD_v/dE_v$ is a dielectric function, and $l_u$, $l_v$ and $l_w$ are the Lam\'{e} coefficients \cite{dolin1961possibility}. When selecting cylindrical coordinate system $(r,\phi,z)$, taking $r$ axis as $u$ axis and considering $z$-axis polarization, these equations are reduced into a model of cylindrical waves which are studied in Refs. \cite{2010Petrov190404,2010Xiong57602,2010Kudrin537,2015Xiong11071,2011Xiong43841,2011Eskin67602,
2011Xiong63845,2012Petrov55202,2012Xiong16602,2012Xiong16606,2013Chen35202,2016Ranjbar19}. For convenience, we introduce the dimensionless variables $\rho=u/u_0$, $\tau=ct/u_0$, $ E=E_v/\text{(N/C)}$, and $ H=\mu_0cH_w/\text{(N/C)}$, where $u_0$ is a constant associated with the characteristic spatial scale, $c$ is the speed of light in vacuum, and (N/C) is the unit of electric field strength in the international system of units. Then we can get the dimensionless Maxwell equations as follows:
\begin{equation}
\begin{aligned}
g_1(\rho)\frac{\partial  E(\rho,\tau)}{\partial\rho}+g_2(\rho) E(\rho,\tau)&=\frac{\partial H(\rho,\tau)}{\partial\tau},
\\
g_3(\rho)\frac{\partial  H(\rho,\tau)}{\partial\rho}+g_4(\rho) H(\rho,\tau)&=\epsilon(\rho, E)
\frac{\partial  E(\rho,\tau)}{\partial\tau},
\end{aligned}
\label{eq:eps}
\end{equation}
where $g_1(\rho)$, $g_2(\rho)$, $g_3(\rho)$ and $g_4(\rho)$ are the dimensionless coefficients. Hereafter, dielectric function $\epsilon(\rho, E)$ is used to represent the media.  Define
\begin{equation*}
\begin{aligned}
g_5(\rho)&=\exp\left[\int(g_1(\rho))^{-1}g_2(\rho)d{\rho}\right],
\\
g_6(\rho)&=\exp\left[\int(g_3(\rho))^{-1}g_4(\rho)d{\rho}\right],
\\
f(\rho)&=\frac{g_1(\rho)[g_6(\rho)]^2}{g_3(\rho)[g_5(\rho)]^2},h=\int{\frac{g_5(\rho)}{g_1(\rho)g_6(\rho)}}d{\rho}
=h(\rho),
\end{aligned}
\end{equation*}
where hereafter the integral constants in the indefinite integrals are always set as zero. We use the following ansatz:
\begin{equation}
\begin{aligned}
\xi&=\alpha g_5(\rho) E+\beta h+\gamma,
\eta=\alpha g_6(\rho) H+\beta \tau+\gamma,
\end{aligned}
\label{eq:3}
\end{equation}
where $\alpha$, $\beta$ and $\gamma$ are arbitrary constants ($\beta\neq0$).

\section{Methods}
In contrast with the previous studies starting with a nonlinear system directly \cite{2010Petrov190404,2010Xiong57602,2010Kudrin537,2015Xiong11071,2011Xiong43841,2011Eskin67602,
2011Xiong63845,2012Petrov55202,2012Xiong16602,2012Xiong16606,2013Chen35202,2016Ranjbar19}, our work begins with a linear system and then transforms it into a nonlinear system through a hodograph transformation \cite{1995Pallikaros6459,1998Kingston1597,2003Sophocleous441}. Because there exists some mathematical difficulties in applying the hodograph transformation on nonlinear system directly.
\paragraph{Firstly} We start with a linear system that characterizes the electromagnetic fields in vacuum as follows:
\begin{equation}
\begin{aligned}
g_1(\rho)\frac{\partial E}{\partial\rho}+g_2(\rho) E=\frac{\partial H}{\partial\tau},
g_3(\rho)\frac{\partial H}{\partial\rho}+g_4(\rho) H=\frac{\partial E}{\partial\tau},
\end{aligned}
\label{eq:2}
\end{equation}
which is derived from system~(\ref{eq:eps}) by choosing $\epsilon(\rho, E)=1$. Suppose $ E_{0}= E_{0}(\rho,\tau)$ and $ H_{0}= H_{0}(\rho,\tau)$ is a solution of Eqs.~(\ref{eq:2}). Then by ansatz~(\ref{eq:3}), Eqs.~(\ref{eq:2}) can be reduced to
\begin{equation}
\begin{aligned}
\frac{\partial\xi}{\partial h}&=\frac{\partial\eta}{\partial\tau},
&\frac{\partial\eta}{\partial h}&=f(\rho)\frac{\partial\xi}{\partial\tau}.
\end{aligned}
\label{eq:4}
\end{equation}
A solution of Eqs.~(\ref{eq:4}) is
\begin{equation}
\begin{aligned}
\xi=C_1g_5(\rho) E_{0}+C_2h+C_3,
\eta=C_1g_6(\rho) H_{0}+C_2\tau+C_3,
\end{aligned}
\label{eq:5}
\end{equation}
where $C_1$, $C_2$ and $C_3$ are arbitrary constants.

If the Jacobian of Eqs.~(\ref{eq:4}) $D(\xi,\eta)/D(h,\tau)=0$, then their solutions must have the form $\xi=F_1\bm{(}\tau\pm\int{\sqrt{f(\rho)}}dh\bm{)}$ and $\eta=F_2\bm{(}\tau\pm\int{\sqrt{f(\rho)}}dh\bm{)}$, where $F_1$ and $F_2$ are arbitrary functions which satisfy $F_2'\bm{(}\tau\pm\int{\sqrt{f(\rho)}}dh\bm{)}=\pm\sqrt{f(\rho)}F_1'\bm{(}\tau\pm\int{\sqrt{f(\rho)}}dh\bm{)}$. (Here the prime denotes the derivative.) In particular, $\sqrt{f(\rho)}=\pm F_2'/F_1'=F\bm{(}\tau\pm\int{\sqrt{f(\rho)}}dh\bm{)}$ holds only if $f(\rho)$ is constant. Since our intention is transforming Eqs.~(\ref{eq:4}) into some nonlinear forms, we do not need to consider this trivial case. When the Jacobian $D(\xi,\eta)/D(h,\tau)\neq0$, the hodograph transformation is applicable, and hence we can view $\xi$ and $\eta$ as independent variables and obtain
\begin{equation}
\begin{aligned}
\frac{\partial h}{\partial\xi}&=\frac{\partial\tau}{\partial\eta},
&\frac{\partial \tau}{\partial\xi}&=f(\rho)\frac{\partial h}{\partial\eta}.
\end{aligned}
\label{eq:6}
\end{equation}
Here formula~(\ref{eq:5}) is still valid.

Now we look at system~(\ref{eq:eps}) for the case
\begin{equation}
\begin{aligned}
\epsilon(\rho, E)=f(\sigma)/f(\rho),
\end{aligned}
\label{eq:7}
\end{equation}
where $\sigma=h^{-1}\bm{(}\alpha g_5(\rho) E+ \beta h+\gamma\bm{)}=h^{-1}(\xi)$ and $h^{-1}(\xi)$ is the inverse function of $h(\xi)$. Through ansatz~(\ref{eq:3}) we get
\begin{equation*}
\begin{aligned}
\frac{\partial\xi}{\partial h}&=\frac{\partial\eta}{\partial\tau},
&\frac{\partial\eta}{\partial h}&=f(\sigma)\frac{\partial\xi}{\partial\tau},
\end{aligned}
\end{equation*}
whose dependent variables ($\xi$ and $\eta$) and independent variables ($h$ and $\tau$) share the same relation as nonlinear system~(\ref{eq:6}). By analogy to formula~(\ref{eq:5}) it yields
\begin{equation}
\begin{aligned}
h=&C_1g_5(\sigma) E_{0}(\sigma,\eta)+C_2\xi+C_3,
\\
\tau=&C_1g_6(\sigma) H_{0}(\sigma,\eta)+C_2\eta+C_3.
\end{aligned}
\label{eq:8}
\end{equation}
Expressing $ E$ and $ H$ from ansatz~(\ref{eq:3}), using formula~(\ref{eq:8}) and setting $C_1=-\alpha/\beta$, $C_2=1/\beta$ and $C_3=-\gamma/\beta$, we obtain
\begin{equation}
\begin{aligned}
 E(\rho,\tau)=\frac{g_5(\sigma)}{g_5(\rho)} E_{0}(\sigma,\eta),
 H(\rho,\tau)=\frac{g_6(\sigma)}{g_6(\rho)} H_{0}(\sigma,\eta).
\end{aligned}
\label{eq:9}
\end{equation}
These expressions give an exact solution of Maxwell equations~(\ref{eq:eps}) in inhomogeneous nonlinear and nondispersive media~(\ref{eq:7}).

Typically, the Maxwell equations have a set of boundary and initial conditions. For the preassigned conditions of Eqs.~(\ref{eq:eps}) and (\ref{eq:7}), it is not difficult to get the corresponding conditions of linear Eqs.~(\ref{eq:2}). For example, if Eqs.~(\ref{eq:eps}) and (\ref{eq:7}) meet the initial conditions
\begin{equation*}
\begin{aligned}
 E(\rho,0)=\varphi_1(\rho),
 H(\rho,0)=\varphi_2(\rho),
\end{aligned}
\end{equation*}
then the solution of Eqs.~(\ref{eq:2}) must meet the generalized initial conditions
\begin{equation*}
\begin{aligned}
 E_0(\sigma_0,\eta_0)=\frac{\varphi_1(\rho)g_5(\rho)}{g_5(\sigma_0)},
 H_0(\sigma_0,\eta_0)=\frac{\varphi_2(\rho)g_6(\rho)}{g_6(\sigma_0)},
\end{aligned}
\end{equation*}
 where $\sigma_0=h^{-1}\bm{(}\alpha g_5(\rho)\varphi_1(\rho)+ \beta h(\rho)+\gamma\bm{)}$ and $\eta_0=\alpha g_6(\rho)\varphi_2(\rho)+\gamma$.

\paragraph{Secondly} Substituting the following ansatz into linear system~(\ref{eq:2}):
\begin{equation*}
\begin{aligned}
\bar{h}&=\int{\frac{g_6(\rho)}{g_3(\rho)g_5(\rho)}}d{\rho}=\bar{h}(\rho),
\\
\bar{\xi}&=\alpha g_5(\rho) E+\beta \tau+\gamma,
\bar{\eta}=\alpha g_6(\rho) H+\beta \bar{h}+\gamma,
\end{aligned}
\end{equation*}
we get
\begin{equation}
\begin{aligned}
\frac{\partial\bar{\xi}}{\partial\bar{h}}&=\frac{1}{f(\rho)}\frac{\partial\bar{\eta}}{\partial\tau},
&\frac{\partial\bar{\eta}}{\partial\bar{h}}&=\frac{\partial\bar{\xi}}{\partial\tau}.
\end{aligned}
\label{eq:12}
\end{equation}
Similarly, assuming $ E_{1}= E_{1}(\rho,\tau)$ and $ H_{1}= H_{1}(\rho,\tau)$ is a solution of system~(\ref{eq:2}), then a solution of system~(\ref{eq:12}) is
\begin{equation*}
\begin{aligned}
\bar{\xi}=-\frac{\alpha}{\beta}g_5(\rho) E_1+\frac{1}{\beta}\tau-\frac{\gamma}{\beta},
\bar{\eta}=-\frac{\alpha}{\beta}g_6(\rho) H_1+\frac{1}{\beta}\bar{h}-\frac{\gamma}{\beta}.
\end{aligned}
\end{equation*}
By applying the hodograph transformation to linear system~(\ref{eq:12}), we get the nonlinear system
\begin{equation}
\begin{aligned}
\frac{\partial\tau}{\partial\bar{\eta}}&=\frac{1}{f(\rho)}\frac{\partial\bar{h}}{\partial\bar{\xi}},
&\frac{\partial\tau}{\partial\bar{\xi}}&=\frac{\partial\bar{h}}{\partial\bar{\eta}}.
\end{aligned}
\label{eq:13}
\end{equation}

When the function $\epsilon$ is chosen such that
\begin{equation}
\begin{aligned}
\epsilon(\rho, E)=1/[f(\rho)f(\bar{\sigma})],
\end{aligned}
\label{eq:14}
\end{equation}
where $\bar{\sigma}=\bar{h}^{-1}\bm{(}\alpha g_5(\rho) E+ \beta h+\gamma\bm{)}=\bar{h}^{-1}(\xi)$, then system~(\ref{eq:eps}) share the same form as system~(\ref{eq:13}): $\partial_{h}\eta=[f(\bar{\sigma})]^{-1}\partial_{\tau}\xi$, $\partial_{h}\xi=\partial_{\tau}\eta$, by ansatz~(\ref{eq:3}). Hence $\tau$ and $h$ satisfy
\begin{equation}
\begin{aligned}
\tau=&-\frac{\alpha}{\beta}g_5(\bar{\sigma}) E_1(\bar{\sigma},\eta)+\frac{1}{\beta}\eta-\frac{\gamma}{\beta},
\\
h=&-\frac{\alpha}{\beta}g_6(\bar{\sigma}) H_1(\bar{\sigma},\eta)+\frac{1}{\beta}\xi-\frac{\gamma}{\beta}.
\end{aligned}
\label{eq:15}
\end{equation}
Substituting ansatz~(\ref{eq:3}) into formula~(\ref{eq:15}), we can write an exact solution of the Maxwell equations in inhomogeneous nonlinear and nondispersive media~(\ref{eq:14}) as:
\begin{equation}
\begin{aligned}
 E(\rho,\tau)=\frac{g_6(\bar{\sigma})}{g_5(\rho)} H_1(\bar{\sigma},\eta),
 H(\rho,\tau)=\frac{g_5(\bar{\sigma})}{g_6(\rho)} E_1(\bar{\sigma},\eta).
\end{aligned}
\label{eq:16}
\end{equation}

\paragraph{Thirdly} We consider a more generic linear system arising from~(\ref{eq:eps}), in which $\epsilon$ satisfies:
\begin{equation}
\begin{aligned}
\epsilon(\rho, E)=g\bm{(}h(\rho)\bm{)}/f(\rho),
\end{aligned}
\label{eq:17}
\end{equation}
with $g$ being an arbitrary function of $h(\rho)$. If we define
\begin{equation*}
\begin{aligned}
\tilde{g}_3(\rho)=g_3(\rho)f(\rho)/g\bm{(}h(\rho)\bm{)},
\tilde{g}_4(\rho)=g_4(\rho)f(\rho)/g\bm{(}h(\rho)\bm{)},
\end{aligned}
\end{equation*}
then the linear system becomes
\begin{equation}
\begin{aligned}
g_1(\rho)\frac{\partial E}{\partial\rho}+g_2(\rho) E=\frac{\partial H}{\partial\tau},
\tilde{g}_3(\rho)\frac{\partial H}{\partial\rho}+\tilde{g}_4(\rho) H=
\frac{\partial E}{\partial\tau},
\end{aligned}
\label{eq:18}
\end{equation}
which is a case we have just discussed. Hence an exact solution of the following nonlinear system
\begin{equation*}
\begin{aligned}
g_1(\rho)\frac{\partial E}{\partial\rho}+g_2(\rho) E&=\frac{\partial H}{\partial\tau},
\\
\tilde{g}_3(\rho)\frac{\partial H}{\partial\rho}+\tilde{g}_4(\rho) H&=\frac{g(\xi)}
{g\bm{(}h(\rho)\bm{)}}\frac{\partial E}{\partial\tau},
\end{aligned}
\end{equation*}
is given by formula~(\ref{eq:9}), where $ E_0(\rho,\tau)$ and $ H_0(\rho,\tau)$ represent a solution of system~(\ref{eq:18}). In other words, if $ E_0$ and $ H_0$ is a solution of Maxwell equations~(\ref{eq:eps}) in linear media~(\ref{eq:17}), then an exact solution of Eqs.~(\ref{eq:eps}) in the inhomogeneous nonlinear and nondispersive media
\begin{equation}
\begin{aligned}
\epsilon(\rho, E)=g(\xi)/f(\rho)=g\bm{(}\alpha g_5(\rho) E+\beta h+\gamma\bm{)}/f(\rho),
\end{aligned}
\label{eq:19}
\end{equation}
is Eqs.~(\ref{eq:9}).

Similarly, defining $G(\rho)=\int g(\rho)d\rho$ and $\tilde{\sigma}=h^{-1}\bm{(}G(\xi)\bm{)}$, and assuming $ E_{1}= E_{1}(\rho,\tau)$ and $ H_{1}= H_{1}(\rho,\tau)$ being a solution of Eqs.~(\ref{eq:eps}) in the linear media
\begin{equation}
\begin{aligned}
\epsilon(\rho, E)=1/[f(\rho)g\bm{(}G^{-1}(h(\rho))\bm{)}],
\end{aligned}
\label{eq:20}
\end{equation}
we can also find an exact solution of the Maxwell equations in media~(\ref{eq:19}) according to formula~(\ref{eq:16}):
\begin{equation}
\begin{aligned}
 E(\rho,\tau)=\frac{g_6(\tilde{\sigma})}{g_5(\rho)} H_1(\tilde{\sigma},\eta),
 H(\rho,\tau)=\frac{g_5(\tilde{\sigma})}{g_6(\rho)} E_1(\tilde{\sigma},\eta).
\end{aligned}
\label{eq:21}
\end{equation}

\section{Results and discussion}
\subsection{Results}
Up to now, we have shown that from the solution of the Maxwell equations~(\ref{eq:eps}) in linear and homogeneous media $\epsilon(\rho, E)=1$, one can obtain the exact solution in nonlinear media~(\ref{eq:7}) through formula~(\ref{eq:9}) and in nonlinear media~(\ref{eq:14}) through formula~(\ref{eq:16}). And the exact solution of nonlinear system~(\ref{eq:eps}) and (\ref{eq:19}) can be got from linear media~(\ref{eq:17}) by formula~(\ref{eq:9}), or from linear media~(\ref{eq:20}) by formula~(\ref{eq:21}). Since $g$ in media~(\ref{eq:19}) is an arbitrary function, we have obtained the exact solutions of electromagnetic wave in arbitrarily nonlinear and particularly inhomogeneous media without dispersion, using the solutions of either of the corresponding two linear systems. For media~(\ref{eq:19}), $\alpha$ and $\beta$ embody the intensity of the nonlinear and inhomogeneous factor, respectively. The larger value $|\alpha|$ means stronger nonlinearity. In practice, by using different substrate materials and designing different cellular architectures, the metamaterial opens a door to realize all possible material properties \cite{2014Lapine1093,2009Cui,2000Pendry3966,1992Sipe1614,2009Boyd1074,2006Schurig977,2006Pendry1780}. For Eqs.~(\ref{eq:eps}) and media~(\ref{eq:19}) with the given definite-solution conditions, we can choose a simpler linear media from (\ref{eq:17}) and (\ref{eq:20}), as formulas~(\ref{eq:9}) and (\ref{eq:21}) return the same result. Taking $\alpha=0$, $\beta=1$ and $\gamma=0$, the following equations represent the relationship between linear media (\ref{eq:17}) and linear media (\ref{eq:20}):
\begin{equation}
\begin{aligned}
E_0(\rho,\tau)&=\frac{g_6\bm{(}h^{-1}(G(h))\bm{)}}{g_5(\rho)}H_1\bm{(}h^{-1}(G(h)),\tau\bm{)},
\\
H_0(\rho,\tau)&=\frac{g_5\bm{(}h^{-1}(G(h))\bm{)}}{g_6(\rho)}E_1\bm{(}h^{-1}(G(h)),\tau\bm{)},
\end{aligned}
\label{eq:22}
\end{equation}
which can be verified by straightforward differentiation. Taking $\beta=1$ and $\gamma=0$, with the weakening of nonlinearity, in the limit $\alpha\rightarrow0$,  the solution $ E$ and $ H$ is reduced into the solution $ E_0$ and $ H_0$ which corresponds to linear media~(\ref{eq:17}).

When selecting cylindrical coordinate system $(r,\phi,z)$, taking $r$ axis as $u$ axis and considering $z$-axis polarization, we have $g_1(\rho)=1$, $g_2(\rho)=0$, $g_3(\rho)=1$, $g_4(\rho)=\frac{1}{\rho}$, $g_5(\rho)=1$, $g_6(\rho)=\rho$, $h(\rho)=\ln\rho$ and $f(\rho)=\rho^2$. If taking $g$ so that media (\ref{eq:17}) satisfy $g(h(\rho))/f(\rho)=1$, then from formula~(\ref{eq:9}) we get
\begin{equation*}
\begin{aligned}
&E(\rho,\tau)= E_0(\rho^\beta e^{\alpha E},\alpha\rho H+\beta\tau),
 \\
&H(\rho,\tau)=\rho^{\beta-1} e^{\alpha E} H_0(\rho^\beta e^{\alpha E},\alpha\rho H+\beta\tau),
\end{aligned}
\end{equation*}
which represents the exact solution of the Maxwell equations in nonlinear inhomogeneous media $\rho^{2(\beta-1)}\exp(2\alpha E)$. This is the case discussed in Refs. \cite{2010Petrov190404,2010Xiong57602,2010Kudrin537,2015Xiong11071,2011Xiong43841,2011Eskin67602,
2011Xiong63845,2012Petrov55202,2012Xiong16602,2012Xiong16606,2013Chen35202,2016Ranjbar19}.

When selecting cylindrical coordinate system $(r,\phi,z)$, taking $r$ axis as $u$ axis and considering $\phi$-axis polarization, we have $g_1(\rho)=-1$, $g_2(\rho)=-\frac{1}{\rho}$, $g_3(\rho)=-1$, $g_4(\rho)=0$, $g_5(\rho)=\rho$, $g_6(\rho)=1$, $h(\rho)=-\frac{\rho^2}{2}$ and $f(\rho)=\frac{1}{\rho^2}$. In this case, the specific expression of formula~(\ref{eq:9}) is
\begin{equation}
\begin{aligned}
 E(\rho,\tau)=&\frac{\sqrt{\beta\rho^2-2\alpha\rho E}}{\rho}
 E_{0}(\sqrt{\beta\rho^2-2\alpha\rho E},\alpha H+\beta \tau),
 \\
 H(\rho,\tau)=& H_{0}(\sqrt{\beta\rho^2-2\alpha\rho E},\alpha H+\beta \tau),
\end{aligned}
\label{eq:10}
\end{equation}
and the specific expression of formula~(\ref{eq:21}) is
\begin{equation}
\begin{aligned}
 E(\rho,\tau)=&\frac{1}{\rho} H_1(e^{\frac{\beta}{2}\rho^2-\alpha\rho E},\alpha H+\beta \tau),
 \\
 H(\rho,\tau)=&e^{\frac{\beta}{2}\rho^2-\alpha\rho E}
 E_1(e^{\frac{\beta}{2}\rho^2-\alpha\rho E},\alpha H+\beta \tau).
\end{aligned}
\label{eq:11}
\end{equation}
Note that the $\phi$-axis polarized and the $z$-axis polarized \cite{2010Petrov190404,2010Xiong57602} cylindrical waves are asymmetric in nonlinear media.  To the best of our knowledge, the features of $\phi$-axis polarized cylindrical waves in nonlinear media remain unknown. Now, let us discuss some applications of those results.

\subsection{Initial value problem and cylindrical shock waves}
Consider a $\phi$-axis polarized cylindrical wave in vacuum with the following form:
\begin{subequations}
\begin{equation}
\begin{aligned}
-\frac{\partial E}{\partial\rho}-\frac{ E}{\rho}=\frac{\partial H}{\partial\tau},
-\frac{\partial H}{\partial\rho}=\frac{\partial E}{\partial\tau},\rho>0,\label{subeq:23a}
\end{aligned}
\end{equation}
\begin{equation}
\begin{aligned}
E(\rho,0)=0,
H(\rho,0)=\varphi(\rho)=1/\sqrt{1+\rho^2},\label{subeq:23b}
\end{aligned}
\end{equation}
\label{eq:23}
\end{subequations}
applying the Hankel transform, we can represent the solution as
\begin{equation}
\begin{aligned}
 E_1(\rho,\tau)&=\int_0^\infty sJ_1(s\rho)\Psi(s)\sin(s\tau)ds,
\\
 H_1(\rho,\tau)&=\int_0^\infty sJ_0(s\rho)\Psi(s)\cos(s\tau)ds,
\end{aligned}
\label{eq:24}
\end{equation}
where $J_\nu$ represents the first kind of the order $\nu$ of the Bessel function \cite{2016Aleahmad2047} and $\Psi(s)=\int_0^\infty\rho J_0(\rho s)\varphi(\rho)d\rho$. According to Eq.~(\ref{eq:7}) and Eq.~(\ref{eq:14}), the corresponding two nonlinear media are $\epsilon(\rho, E)=\frac{\rho}{\beta\rho-2\alpha E}$ and $\epsilon(\rho, E)=\rho^2e^{\beta\rho^2-2\alpha\rho E}$. As mentioned earlier, $\alpha$ and $\beta$ embody the intensity of the nonlinear and inhomogeneous factors, respectively. Ferroelectric bulk materials, ferroelectric thin films and superlattices are typical examples of such a medium
\cite{2006Chen4171,2016Zubko524,2004Fong1650}. Note that ferroelectric materials are known to be of great interest for many promising applications \cite{2005Dawber1083}. After substituting Eqs.~(\ref{eq:24}) into formula~(\ref{eq:11}), the exact solution of the initial value problem
\begin{equation*}
\begin{aligned}
&-\frac{\partial E}{\partial\rho}-\frac{ E}{\rho}=\frac{\partial H}{\partial\tau},
-\frac{\partial H}{\partial\rho}=
\rho^2e^{\beta\rho^2-2\alpha\rho E}\frac{\partial E}{\partial\tau},
\rho>0,
\\
&0<\beta,  E(\rho,0)=\varphi(e^{\frac{\beta}{2}\rho^2-\alpha\rho E(\rho,0)})/\rho,
 H(\rho,0)=0,
\end{aligned}
\end{equation*}
can be written as
\begin{equation}
\begin{aligned}
 E(\rho,\tau)=&\int_0^\infty \frac{s}{\rho}J_0(se^{\frac{\beta}{2}\rho^2-\alpha\rho E})\Psi(s)
\cos[s(\alpha H+\beta\tau)]ds
\\
=&\frac{1}{\rho}\text{Re}\{\frac{1}{\sqrt{(1-i(\alpha H+\beta\tau))^2+
 e^{\beta\rho^2-2\alpha\rho E}}}\},
\\
 H(\rho,\tau)=&e^{\frac{\beta}{2}\rho^2-\alpha\rho E} \int_0^\infty sJ_1(se^{\frac{\beta}{2}\rho^2-\alpha\rho E})\Psi(s)
\sin[s(\alpha H+\beta\tau)]ds
\\
=&\text{Re}\{\frac{i+(\alpha H+\beta\tau)}
{\sqrt{(1-i(\alpha H+\beta\tau))^2+e^{\beta\rho^2-2\alpha\rho E}}}\}.
\end{aligned}
\label{eq:25}
\end{equation}
Here $i$ is the imaginary unit. Note that the function $\varphi$ describing the initial state of the magnetic field in vacuum can be chosen arbitrarily. One should only ensure the convergence of the integrals in Eqs.~(\ref{eq:24}).

\begin{figure}[!htbp]
\includegraphics[scale=0.55]{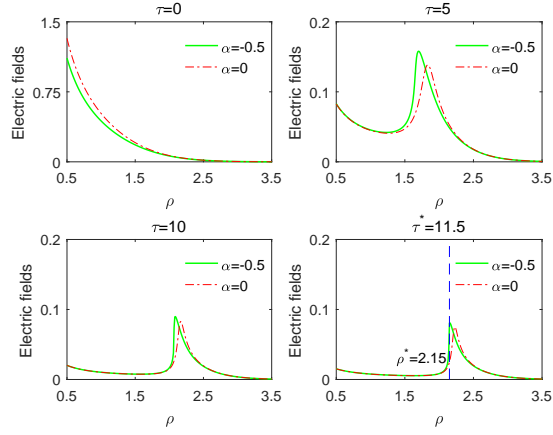}
\caption{\label{fig:epsart} Linear ($\alpha=0$, dash-dotted curves) and nonlinear ($\alpha=-0.5$, solid curves) electric fields of cylindrical wave~(\ref{eq:25})  as function of the coordinate $\rho$ at various times.}
\end{figure}
Figure~\ref{fig:epsart} shows the results of numerical calculation of $ E$ using formula~(\ref{eq:25}) in the cases with $\alpha=0$, $\beta=1$ and $\alpha=-0.5$, $\beta=1$. At various $\tau$, the electric fields as function of the coordinate $\rho$ in linear inhomogeneous media $\epsilon(\rho, E)=\rho^2e^{\rho^2}$ and nonlinear inhomogeneous media $\epsilon(\rho, E)=\rho^2e^{\rho^2+\rho E}$ are shown by the dash-dotted and solid curves, respectively. It is worth mentioning that the velocity of the peak of the nonlinear electric field is slower than that of the trailing edge, and thus leads to the trailing edge becomes increasingly steeper with increasing $\tau$, which is called self-steepening \cite{1965Rosen539, 1967DeMartini312} and eventually creates a cylindrical electromagnetic shock wave at a point $\rho=\rho^*$ at time $\tau=\tau^*$. Such phenomena cannot be described satisfactorily by perturbation theory. In the absence of dispersion, the solution described by Eqs.~(\ref{eq:25}) becomes unsuitable when $\tau>\tau^*$ because multiple different values of $ E$ and $ H$ satisfying Eqs.~(\ref{eq:25}) correspond to the same value of $\rho$, leading to discontinuities of wave components.

\subsection{Boundary value problem and free nonlinear oscillations}
Now we consider the situation that the linear wave Eq.~(\ref{subeq:23a}) satisfies boundary conditions
\begin{equation*}
\begin{aligned}
1<a\leq\rho\leq b, H(a,\tau)=0, H(b,\tau)=0,
\end{aligned}
\end{equation*}
which describes the electromagnetic oscillations of cylindrical waves.  We can obtain the solution of such a system by the method of variable separation:
\begin{equation*}
\begin{aligned}
E_1(\rho,\tau)&=A[Y_0(\omega_na)J_1(\omega_n\rho)-J_0(\omega_na)Y_1(\omega_n\rho)]\sin(\omega_n\tau),
\\
H_1(\rho,\tau)&=A[Y_0(\omega_na)J_0(\omega_n\rho)-J_0(\omega_na)Y_0(\omega_n\rho)]\cos(\omega_n\tau),
\end{aligned}
\end{equation*}
where $J_\nu$ and $Y_\nu$ represent the first and the second kind of the order $\nu$ of the Bessel function respectively, $\omega_n$ is the $n$th positive root of the equation $J_0(a\omega)Y_0(b\omega)-J_0(b\omega)Y_0(a\omega)=0$, and $A$ is an amplitude factor. According to formula~(\ref{eq:11}), the exact solution of the following boundary value problem
\begin{equation*}
\begin{aligned}
&-\frac{\partial E}{\partial\rho}-\frac{ E}{\rho}=\frac{\partial H}{\partial\tau},
-\frac{\partial H}{\partial\rho}=
\rho^2e^{\beta\rho^2-2\alpha\rho E}\frac{\partial E}{\partial\tau},
0<\beta,
\\
&(\frac{2}{\beta}\ln a)^\frac{1}{2}\leq\rho\leq(\frac{2}{\beta}\ln b)^\frac{1}{2},  E((\frac{2}{\beta}\ln a)^\frac{1}{2},0)=E((\frac{2}{\beta}\ln b)^\frac{1}{2},0)=0,
\end{aligned}
\end{equation*}
which describes the corresponding nonlinear oscillations of cylindrical electromagnetic waves, can be written as
\begin{equation}
\begin{aligned}
 E=&[Y_0(\omega_na)J_0(\omega_ne^{\frac{\beta}{2}\rho^2-\alpha\rho E})
 -J_0(\omega_na)Y_0(\omega_ne^{\frac{\beta}{2}\rho^2-\alpha\rho E})]
 \\
 &\times\frac{A}{\rho}\cos[\omega_n(\alpha H+\beta\tau)],
\\
 H=&[Y_0(\omega_na)J_1(\omega_ne^{\frac{\beta}{2}\rho^2-\alpha\rho E})
 -J_0(\omega_na)Y_1(\omega_ne^{\frac{\beta}{2}\rho^2-\alpha\rho E})]
 \\
 &\times Ae^{\frac{\beta}{2}\rho^2-\alpha\rho E}\sin[\omega_n(\alpha H+\beta\tau)].
\end{aligned}
\label{eq:29}
\end{equation}

\begin{figure}[!htbp]
\includegraphics[scale=0.55]{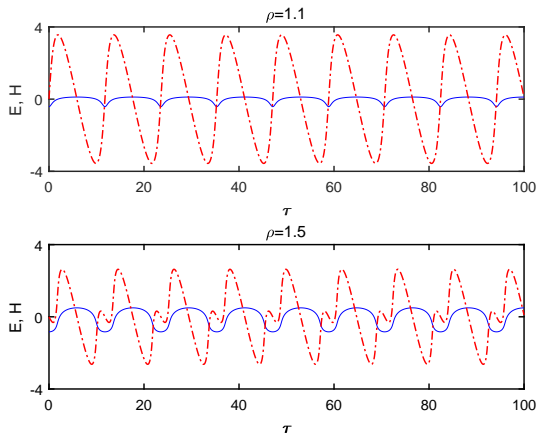}
\caption{ Electric (solid curves) and magnetic (dash-dotted curves) fields of cylindrical wave~(\ref{eq:29}) as functions of $\tau$ in the $n=1$ mode by taking $\alpha=0.3$, $\beta=1$, $a=e^{0.5}$, $b=e^2$ and $A=3$. \label{fig:2}}
\end{figure}
Figure~\ref{fig:2} shows the results of numerical calculation of the fields using solution~(\ref{eq:29}) in the case with $\alpha=0.3$, $\beta=1$, $a=e^{0.5}$, $b=e^2$ and $A=3$ in the $n=1$ mode ($\omega_1\approx0.5336$) at two points $\rho=1.1$ and $\rho=1.5$. We see that the electric field and the magnetic field fluctuate with the same frequency $\omega_n/(2\pi)$ at $\rho=1.1$. However, the electric field varies at frequency $\omega_n/(2\pi)$ while the magnetic field at the second harmonic $\omega_n/\pi$ at $\rho=1.5$, i.e., the second harmonic component which is one of the most intensively studied effects in nonlinear optics is pronounced.

Similar to Ref. \cite{2010Petrov190404}, it should be noted that the electric field $E$ and magnetic field $H$ determined by~(\ref{eq:29}) become ambiguous when the mode $n\geq n^*$, where threshold $n^*$ is an integer depending on the parameter $\alpha$, $\beta$, $a$, $b$ and $A$ (e.g., $n^*=2$ for $\alpha=0.1$, $\beta=1$, $a=1$, $b=2$ and $A=1$). Due to such phenomena are not physically admissible, the cylindrical standing waves solution in form~(\ref{eq:29}) obtained without allowance for dispersion ceases to be suitable for the higher modes.

\subsection{Superposition principle and sum- and difference-frequency generation}
Consider two $\phi$-axis polarized cylindrical waves with frequencies $\varpi_1$ and $\varpi_2$ propagating in the semi infinite inhomogeneous and nonlinear medium $\epsilon(\rho, E)=\frac{\rho}{\beta\rho-2\alpha E}$:
\begin{equation}
\begin{aligned}
&-\frac{\partial E}{\partial\rho}-\frac{ E}{\rho}=\frac{\partial H}{\partial\tau},
-\frac{\partial H}{\partial\rho}=\frac{\rho}{(\beta\rho-2\alpha E)}
\frac{\partial E}{\partial\tau},
\\
&\beta>0, 0<a\leq\rho, E(a,\tau)=0.
\end{aligned}
\label{eq:26}
\end{equation}
According to Eq.~(\ref{eq:17}), one of the corresponding linear system is Eq.~(\ref{subeq:23a}), with the boundary conditions
\begin{equation*}
\begin{aligned}
0<\sqrt{\beta}a\leq\rho, {E}_0(\sqrt{\beta}a,\tau)=0.
\end{aligned}
\end{equation*}
By the superposition principle, we get the solution of the linear problem as follows:
 \begin{equation*}
\begin{aligned}
E_0(\rho,\tau)=&\sum_{i=1}^2A_i[Y_1(\varpi_i\sqrt{\beta}a)J_1(\varpi_i\rho)
\\&-J_1(\varpi_i\sqrt{\beta}a)Y_1(\varpi_i\rho)]\sin(\varpi_i\tau),
\\
H_0(\rho,\tau)=&\sum_{i=1}^2A_i[Y_1(\varpi_i\sqrt{\beta}a)J_0(\varpi_i\rho)
\\&-J_1(\varpi_i\sqrt{\beta}a)Y_0(\varpi_i\rho)]\cos(\varpi_i\tau).
\end{aligned}
\end{equation*}
where $J_\nu$ and $Y_\nu$ represent the first and the second kind of the order $\nu$ of the Bessel function respectively, and $A_1$ and $A_2$ are amplitude factors. According to formula~(\ref{eq:10}), we obtain the solution of Eqs.~(\ref{eq:26}) as follows:
\begin{equation}
\begin{aligned}
E=&\frac{1}{\rho}\sum_{i=1}^2A_i(\beta\rho^2-2\alpha\rho E)^{\frac{1}{2}}
[Y_1(\varpi_i\sqrt{\beta}a)J_1\bm{(}\varpi_i(\beta\rho^2-2\alpha\rho E)^{\frac{1}{2}}\bm{)}
\\&-J_1(\varpi_i\sqrt{\beta}a)Y_1\bm{(}\varpi_i(\beta\rho^2-2\alpha\rho E)^{\frac{1}{2}}\bm{)}]
\sin[\varpi_i(\alpha H+\beta\tau)],
\\
H=&\sum_{i=1}^2A_i[Y_1(\varpi_i\sqrt{\beta}a)J_0\bm{(}\varpi_i(\beta\rho^2-2\alpha\rho E)^{\frac{1}{2}}\bm{)}
\\&-J_1(\varpi_i\sqrt{\beta}a)Y_0\bm{(}\varpi_i(\beta\rho^2-2\alpha\rho E)^{\frac{1}{2}}\bm{)}]
\cos[\varpi_i(\alpha H+\beta\tau)].
\end{aligned}
\label{eq:27}
\end{equation}
Here we should properly choose $A_1$ and $A_2$, to make inequality $2\max(\alpha E/\rho)<\beta$ hold. Such an explicit expression can easily be extended to deal with the problem of any amount of cylindrical electromagnetic waves propagation in nonlinear media, while applying the traditional methods is very difficult \cite{2015Xiong11071}.

\begin{figure}[!htbp]
\includegraphics[scale=0.5]{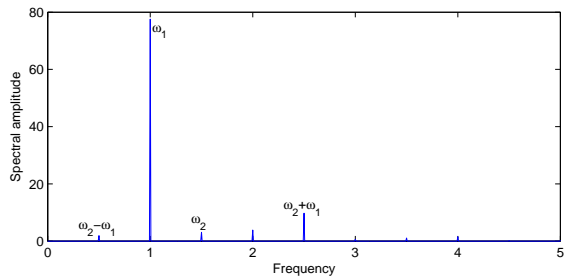}
\caption{Frequency spectrum of the electric field of Eqs.~(\ref{eq:27}) when two cylindrical waves propagate in a semi infinite nonlinear medium. We take $\rho=2$, $\alpha=-1$, $\beta=9$, $a=1$, $A_1=A_2=1$, $\varpi_1=\frac{2\pi}{9}$ and $\varpi_2=\frac{\pi}{3}$ ($\omega_1=\frac{9\varpi_1}{2\pi}$, $\omega_2=\frac{9\varpi_2}{2\pi}$).\label{fig:3}}
\end{figure}
Figure~\ref{fig:3} shows the result of the fast Fourier transform of the electric field determined by Eqs.~(\ref{eq:27}) at $\rho=2$ by taking $\alpha=-1$, $\beta=9$, $a=1$, $A_1=A_2=1$, $\varpi_1=\frac{2\pi}{9}$ and $\varpi_2=\frac{\pi}{3}$ ($\omega_1=\frac{9\varpi_1}{2\pi}$, $\omega_2=\frac{9\varpi_2}{2\pi}$). We can see the sum frequency ($\omega_1+\omega_2$) and the difference frequency ($\omega_2-\omega_1$) which are second-order nonlinear optical processes are pronounced.

\section{Conclusions}
In conclusion, we have revealed the correlation between the linear and nonlinear (1+1)-dimensional Maxwell equations under the orthogonal curvilinear coordinates, and shown how to acquire the exact solutions of the Maxwell equations in nondispersive media with arbitrary nonlinearity and particular inhomogeneity, which have puzzled researchers for a long time. The exact solutions enable us to analytically analyze some nonlinear problems difficultly tackled by traditional methods. As examples, we analyze the initial value problem, boundary value problem and sum- and difference-frequency generation. The results can be applied to studying nonlinear optics and other nonlinear electromagnetic phenomena conveniently and effectively, as well as studying the properties of ferroelectric materials and metamaterials.

%Obviously, our results can be extended to solve the problem of nonlinear magnetization.
% Give an acknowledgement if you have to thank someone, are
% supported by someone or something like that:
\begin{acknowledgments}
We thank Kaijie Chen and Ding Liu for constructive advice.
\end{acknowledgments}

\nocite{*}
%\bibliographystyle{osajnl}
%\bibliography{mynuclear}

\end{document}